\documentclass[]{spie}
 
\usepackage[]{graphicx}
\usepackage{amsfonts}
\usepackage{mathtools}
\usepackage{mathptmx}
\usepackage{subfig}
\usepackage{floatrow}

\usepackage{sidecap}
\title{Spatiotemporal Disentanglement of Arteriovenous Malformations in Digital Subtraction Angiography}

\author{Kathleen Baur\supit{a,b,c}, Xin Xiong\supit{a,b,d}, Erickson Torio\supit{a,b}, Rose Du\supit{a,b}, Parikshit Juvekar\supit{a,b}, Reuben Dorent\supit{a,b}, Alexandra Golby\supit{a,b}, Sarah Frisken\supit{a,b}, Nazim Haouchine\supit{a,b}
\skiplinehalf
\supit{a} Harvard Medical School, Boston, MA, USA; \\
\supit{b} Brigham and Women's Hospital, Boston, MA, USA; \\
\supit{c} Cornell University, Ithaca, NY, USA \\
\supit{d} Columbia University, NYC, NY, USA
}

\begin{document} 
\maketitle 

\begin{abstract}
Although Digital Subtraction Angiography (DSA) is the most important imaging for visualizing cerebrovascular anatomy, its interpretation by clinicians remains difficult. This is particularly true when treating arteriovenous malformations (AVMs), where entangled vasculature connecting arteries and veins needs to be carefully identified.
The presented method aims to enhance DSA image series by highlighting critical information via automatic classification of vessels using a combination of two learning models: An unsupervised machine learning method based on Independent Component Analysis that decomposes the phases of flow and a convolutional neural network that automatically delineates the vessels in image space. 
The proposed method was tested on clinical DSA images series and demonstrated efficient differentiation between arteries and veins that provides a viable solution to enhance visualizations for clinical use. 
\end{abstract}



\section{INTRODUCTION}
\label{sec:intro}
DSA is the premier imaging technique for the visualization of cerebrovascular anatomy before, during, and after neurovascular interventions \cite{spetzler2015comprehensive, frisken2022using}. 
Although it contains essential information regarding hemodynamics and angioarchitecture, clinical analysis of DSA series remains challenging. 
This is particularly true when treating AVMs, where veins and arteries are entangled and need to be carefully identified.
Surgical intervention relies on this knowledge to ensure the preservation of the draining veins until all arterial feeders are interrupted.
Failure to do so may result in post-surgical deficits due to hemorrhage or a potential arterial or venous infarct and may ultimately lead to patient morbidity \cite{spetzler2015comprehensive}.

DSA image series are most commonly enhanced via the application of color-coding techniques to facilitate the identification of malformations and the recognition of these malformations once detected \cite{strother2010parametric,zhang2014color}.
Indeed, colors enhance clinicians' ability to understand complex information and help them perceive slight changes in image intensity, synonymous to changes in blood flow.
DSAs can also provide clinically meaningful measurements of blood flow such as the Maximum Intensity Value and the Time To Peak via the computation of a time-density curve from pixel intensities.
Methods such as phase-based decomposition \cite{lee2017automatic} and optical flow tracking \cite{huang2013quantitative} offer DSA enhancement via color coding for the computation of quantitative vascular flow measurements such as Mean Transient Time, Cerebral Circulation Time and Mean Flow Velocity.
These parameters have been shown to relate to different physiological conditions \cite{strother2010parametric}.

Other methods aim at localizing and classifying vessels or anatomical structures in DSA sequences to extract clinically-relevant information \cite{frisken2022using, momeni2015automatic, podgorvsak2019use}. 
For instance, in order to efficiently detect the location of aneurysms, image-based morphological operations \cite{momeni2015automatic} as well as deep neural networks \cite{podgorvsak2019use} have been successfully applied to angiograms.
In order to automatically score cerebral endovascular treatments, a deep learning approach was proposed in \cite{Su2021} to detect the arterial, capillary, and venous phases. A minimum intensity map was then used to segment the treated region from the healthy one in DSA images.
The authors in \cite{shi2021temporal} showed that it is possible to classify, with accuracy, vessels in angiograms using deep learning.
These methods detect the vessels and AVMs but do not outline the actual geometry of the vessels and would benefit from more robust techniques that automatically segment cerebral vessels in DSA images and can efficiently extract the vascular network from its background \cite{meng2020multiscale, neumann2018convolutional}.

\textbf{Our contribution} is a novel solution to enhance DSA image series of patients afflicted with an AVM.
Given that the nidus is a tangle of vessels connecting arteries to veins, it is crucial to support clinicans in identifying the vasculature surrounding the nidus so that flow is interrupted in the proper order.
Our approach combines an unsupervised machine learning approach that decomposes the phases of flow with a supervised deep learning model for vessel segmentation used to recover temporal information.

\section{METHODS}
\subsection{Temporal Flow Decomposition}
\begin{figure}[t]
    \begin{center}
    \begin{tabular}{c}
        \includegraphics[width=\linewidth]{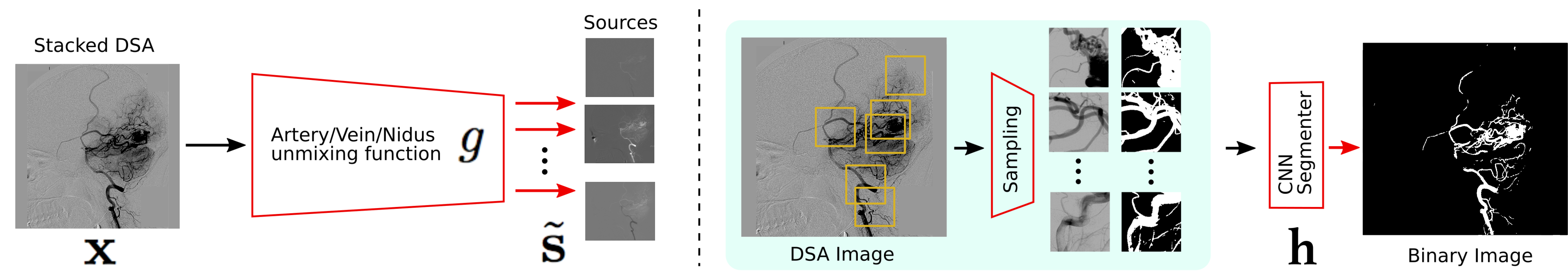}
    \end{tabular}
    \end{center}
    \vspace{-1.5em}
    \caption{Overview of applied techniques: In order to separate phases of vascular flow in a DSA image series $\mathbf{x}$ we use the ICA unmixing function $g$ to decompose $\mathbf{x}$ into arterial, venous, and nidal sources $\tilde{\mathbf{s}}$ and the segmentation model $h$ to produce a binary mask that will be used to build the final visualization.}
    \label{fig:overview}
\end{figure}

Let us denote $\mathbf{x} = [x_1, x_2, \dots, x_d ] \in R^{d\times h\times w}$ a series of $d$ 2D images. 
Let assume that the series $\mathbf{x}$ is generated by $p$ independent components (or independent latent variables) $\mathbf{s} = [s_1, s_2, \dots , s_p] \in \mathbb{R}^{p\times h\times w}$ according to $\mathbf{x} = f(\mathbf{s})$, where $f$ is a mixing function. 
Our objective is to estimate $f^{-1}$ as well as the independent components $[s_1, s_2, \dots , s_p]$ solely based on observations $\mathbf{x}$ following $\tilde{\mathbf{s}} = g(\mathbf{x})$, where $g$ is an unmixing function and $\tilde{\mathbf{s}} = [\tilde{s_1}, \tilde{s_2}, \dots , \tilde{s_p}]$ are the recovered sources.
As there are fewer components (arterial, nidal, and venous phase) than images, i.e. $d \geq p$, and by considering the images to be a linear mixture of three phases, we can predict $g$ and $\tilde{\mathbf{s}}$ using Independent Component Analysis (ICA) (See Fig. \ref{fig:overview}).

There are a variety of algorithms for ICA which, for instance, rely on different measures of independence, including ones minimizing mutual information or maximizing the non-Gaussianity of recovered sources \cite{hyvarinen1999fast, bell1995information}. Methods of the latter category have successfully been applied to cases without AVMs to e.g. extract measurements of blood flow \cite{Haouchine2021Estimation, lee2017automatic}. We first manually trim the DSA image series to remove superfluous frames and borders or to define a region of interest around the malformation to constrain the decomposition to vessels surrounding the nidus. The images are then stacked and vectorized to produce a 1-D signal. The three separate signals produced by the unmixing function $g$ can be transformed back to 2D space to constitute three images corresponding to the feeding arteries, nidus and capillaries, and draining veins.

\subsection{Vessel Classification}
To identify the phase represented in each component and recover temporal information, the spatiotemporal location of vessels in the input series is required. We trained a convolutional neural network in order to segment vasculature and AVMs from the background as illustrated in Fig. \ref{fig:overview}. 

Let us define the training set $T = \{(x_i, l_i)\}_j$ composed of images $x_i$ from DSA image series and their corresponding binary annotations $l_i$, with $j \in |T|$.
We trained a network $h(x_i, \theta_h)$, with $\theta_h$ being the parameters to be learned for network $h$.
This model is used to generate a per-pixel probability map $m$ that will associate for each pixel of an image $x_i$ a probability of being part of the vasculature. 
The model follows standard U-Net architecture \cite{unet} and is trained using gradient descent optimization with Nadam and Dice loss over the parameters $\theta_h$.
In practice, we used small patches to train the network, which permitted us to capture small vessels and augment the dataset. 
The patches were automatically selected using a measure of range entropy to balance the dataset \cite{meng2020multiscale}. We used full-size DSA images for inference.

\subsection{Spatiotemporal re-composition of DSA Series} 
Finally, at run-time, given an image series $\mathbf{x}$ we create a visualization $\mathbf{y}$ through use of the overlap of the phases given by $g(\mathbf{x})$ and the vasculature masks from the application of the segmentation network $h(\mathbf{x},\widehat{\theta_h}))$: 
\begin{equation}
\mathbf{y} \leftarrow  \Gamma_\text{vis} \big(g(\mathbf{x}),  h(\mathbf{x},\hat{\theta_h})\big)
\end{equation}

where $\widehat{\theta_h}$ are the resulting parameters from the training. $\Gamma_\text{vis}$ relies on thresholding of the sources. When the binary sources and vessel masks in temporal order are combined, each phase's maximum intensity throughout time can be used to infer the order of the components. The spatiotemporal location of arteries, capillaries and nidus, and veins can then be used to generate a color-coded image series.

\captionsetup[subfigure]{labelformat=empty}
\begin{figure}[h!]
\centering 
\subfloat[Series A]{\includegraphics[width=0.135\linewidth, height=57pt]{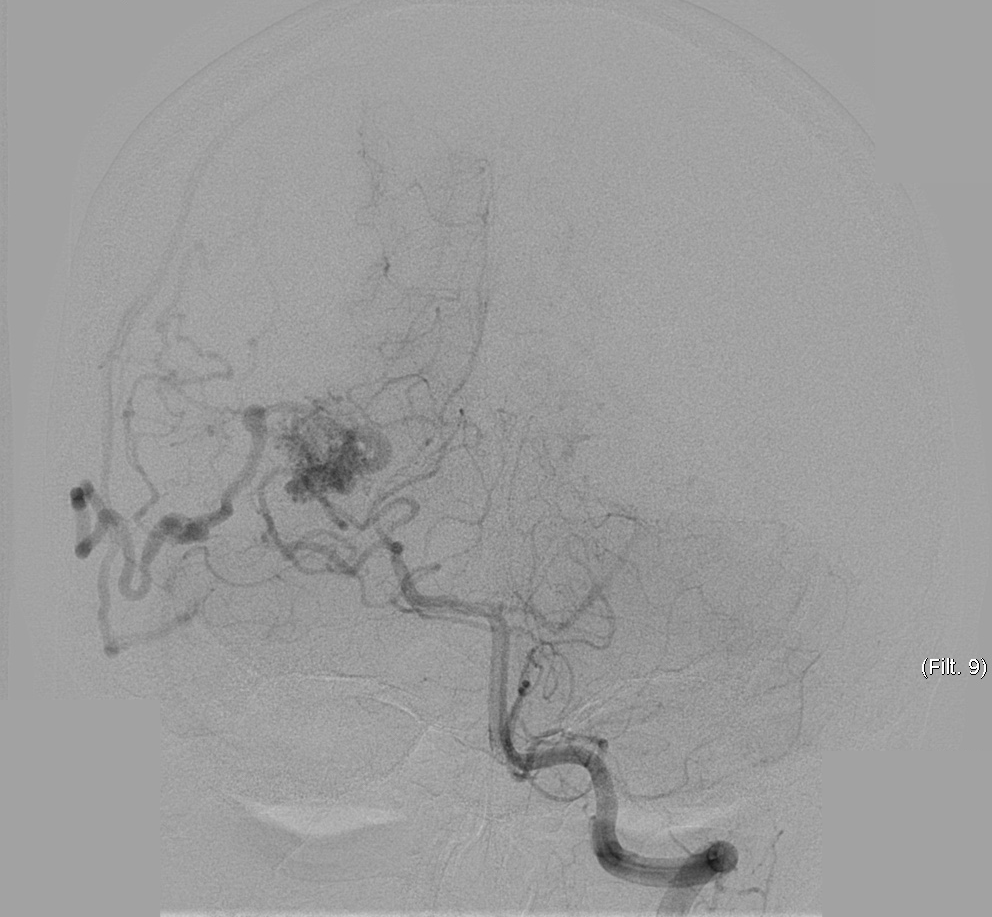}}
\hfill
\subfloat[Phase 1]{\includegraphics[width=0.135\linewidth, height=57pt]{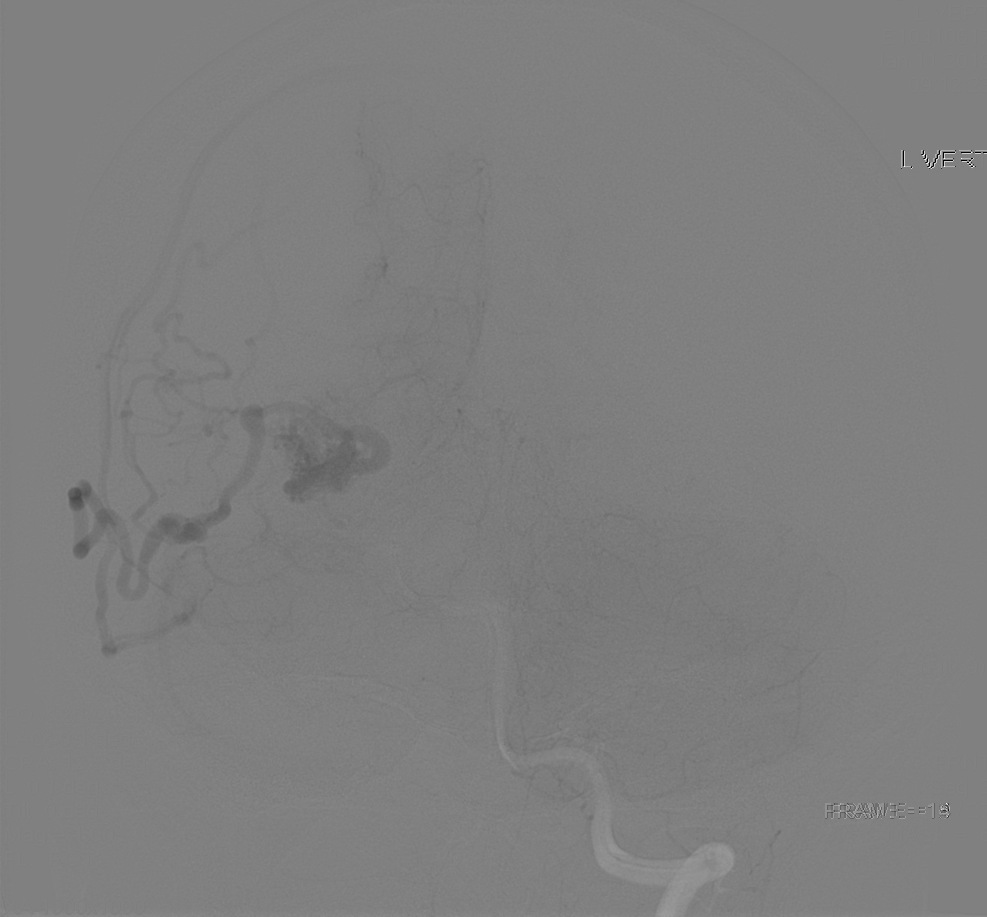}}
\hfill
\subfloat[Phase 2]{\includegraphics[width=0.135\linewidth, height=57pt]{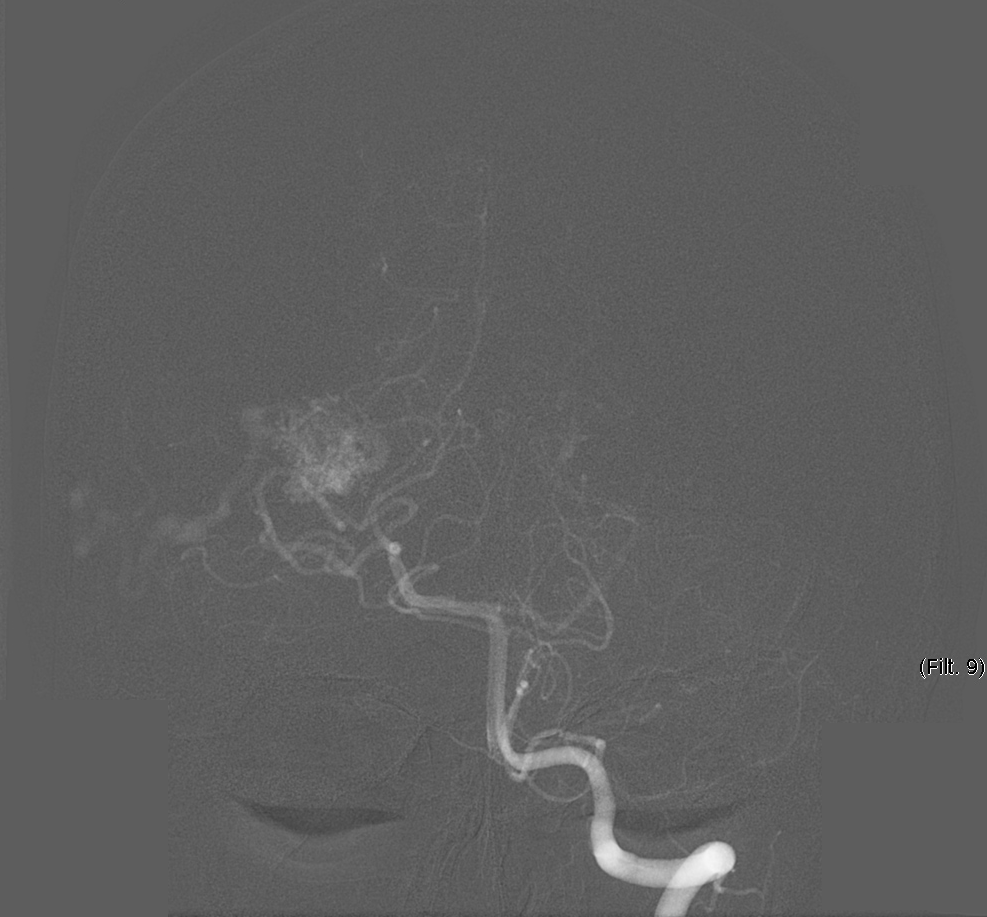}}
\hspace{1.5em}
\subfloat[Series B]{\includegraphics[width=0.135\linewidth, height=57pt]{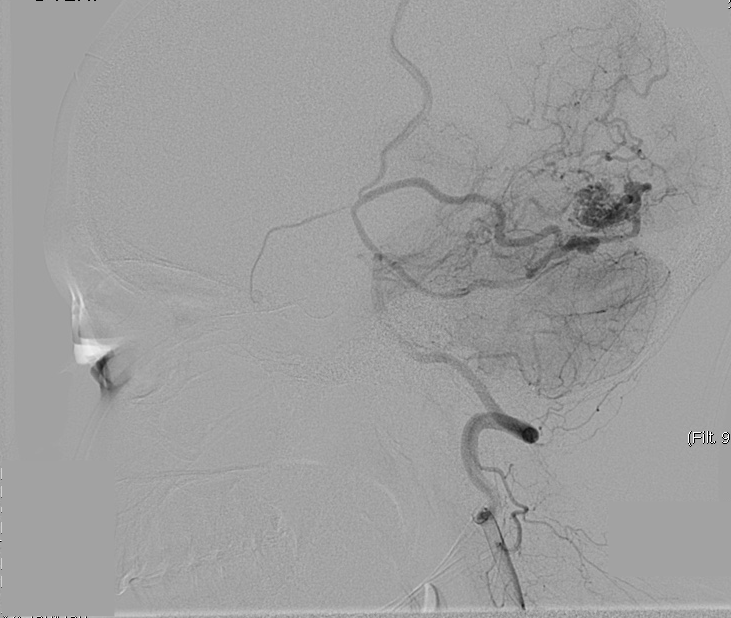}}
\hfill
\subfloat[Phase 1]{\includegraphics[width=0.135\linewidth, height=57pt]{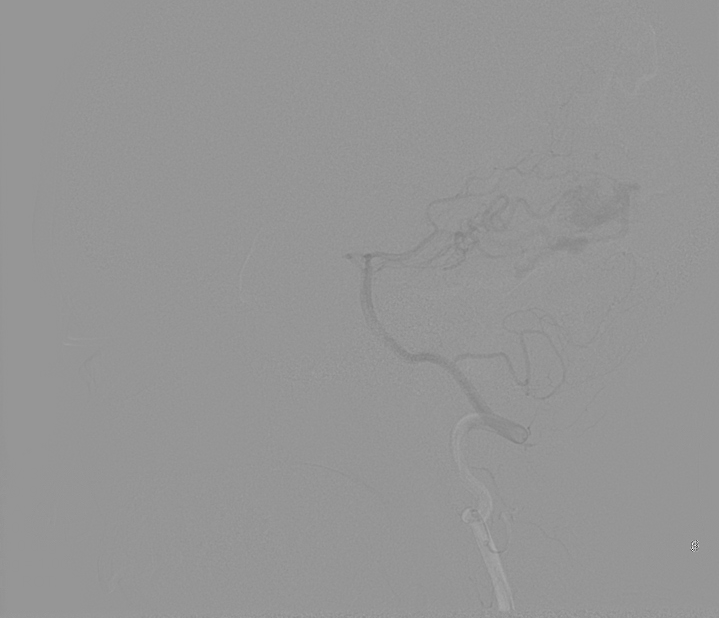}}
\hfill
\subfloat[Phase 2]{\includegraphics[width=0.135\linewidth, height=57pt]{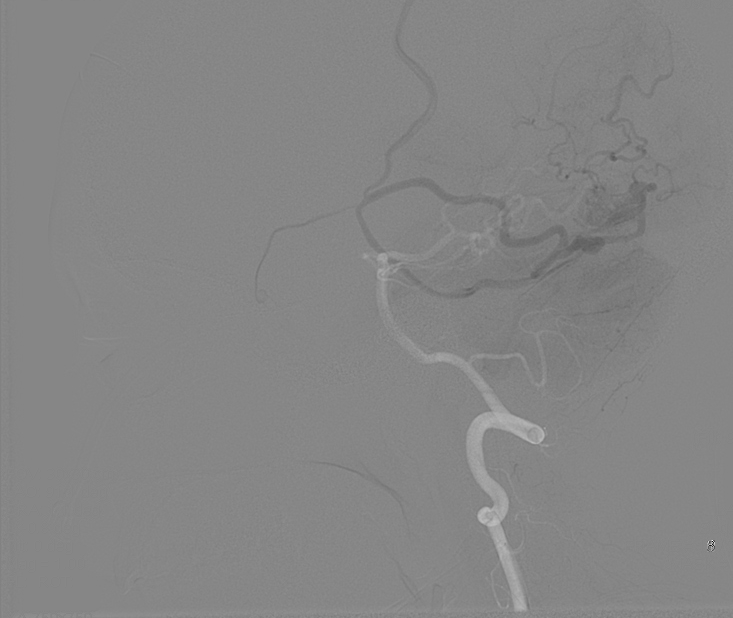}}
\hfill
\subfloat[Phase 3]{\includegraphics[width=0.135\linewidth, height=57pt]{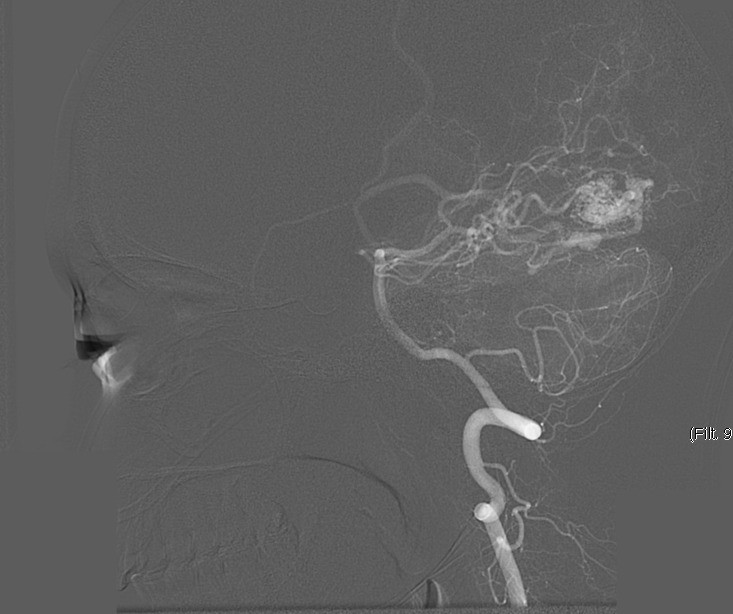}}

\caption{The sources $\tilde{\mathbf{s}} = g(\mathbf{x})$ that result from applying ICA decomposition to two DSA image series (A and B) provide information for phase decomposition. The phases can be visualized as two or three distinct images.}
\label{fig:res}
\end{figure}

\section{RESULTS}

The vessel segmentation model was trained on 864 patches with a fixed size of 256 by 256 pixels which where extracted from DSA images of healthy and diseased vasculature. The patches were manually split into 5 folds for cross validation to prevent data leakage due to a lack of independence between patches from the same DSA series. Gray value, contrast, noise, and brightness variations resulted in 3415 training patches. Furthermore, on the fly affine transformations included horizontal and vertical flips, rotations, zoom, and shear changes to generate a total of 98880 augmented patches. The model was trained over 30 epochs with 100 steps each using a batch size of 32 and a learning rate of 1e-4. Three validation steps per epoch were used for monitoring to reduce the learning rate on plateau and enforce early stopping. MobileNetV2 \cite{sandler2018mobilenetv2} pretrained on ImageNet was used for transfer learning. The network was implemented using the Tensorflow framework (www.tensorflow.org) and trained on an Apple Silicon M2 Pro. The model achieved high scores as measured by a dice score of 0.944, recall of 0.957, and precision of 0.977.

\begin{figure}[h!]
\centering 
\subfloat[]{\includegraphics[width=0.99\linewidth, height=42pt]{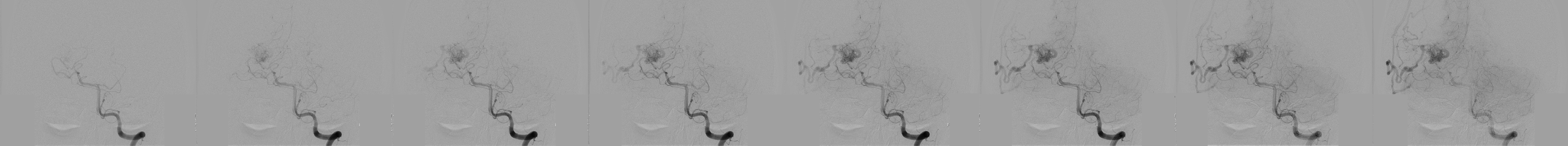}} \\
\vspace{-2.3em}
\subfloat[]{\includegraphics[width=0.99\linewidth, height=42pt]{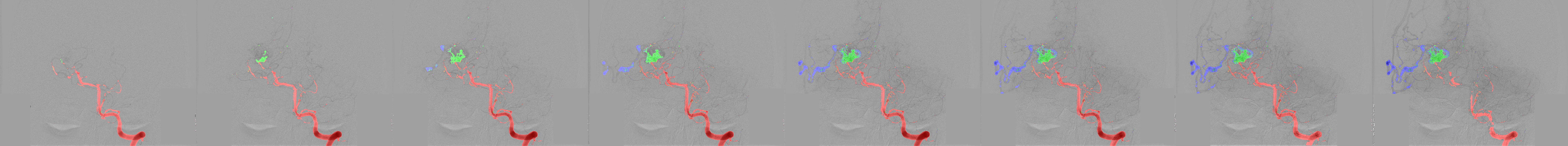}} \\
\vspace{-2.3em}
\subfloat[]{\includegraphics[width=0.99\linewidth, height=42pt]{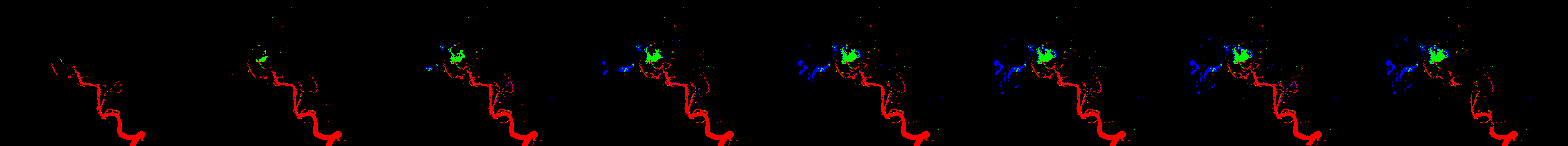}}
\vspace{-0.5em}
\subfloat[]{\includegraphics[width=0.99\linewidth, height=42pt]{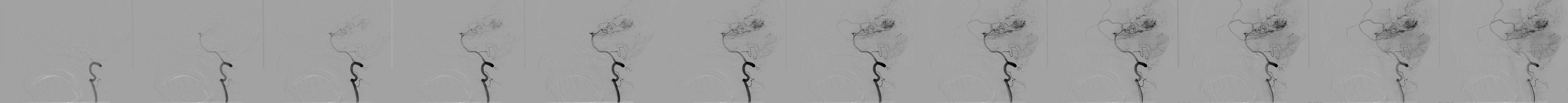}} \\
\vspace{-2.3em}
\subfloat[]{\includegraphics[width=0.99\linewidth, height=42pt]{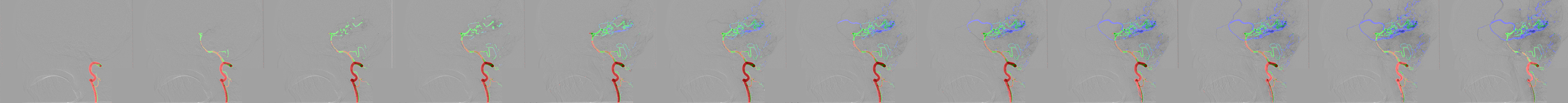}} \\
\vspace{-2.3em}
\subfloat[]{\includegraphics[width=0.99\linewidth, height=42pt]{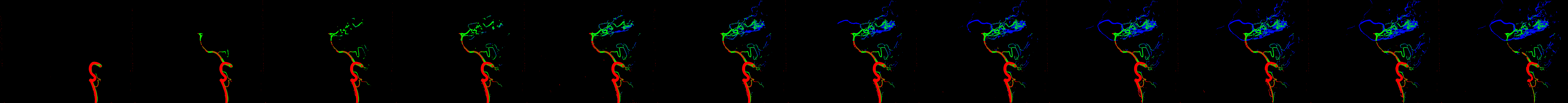}}

\caption{Examples of phase-constrained color-coded DSA image series (A and B) showing the progressive appearance of classified arteries and veins which feed or drain the nidus. Arteries are shown in red, veins in blue, nidus and capillaries in green.}
\label{fig:seq}
\end{figure}

We chose FastICA to define the unmixing function $g$ and applied it on 19 DSA image series of subjects with an AVM.
The series were composed of 12 to 38 images showing the malformations at different time points and were acquired at a low frame rate of 1.5fps to 6fps. 
As shown in Figure~\ref{fig:res}, the function $g$ was capable of decomposing the phases of vascular flow. We also generated phase-constrained color-coded overlays for DSA image series through the combination of the recovered sources and vessel masks, leading to a visualization where contrast flow appears progressively and is classified as artery, nidus and capillary, or vein. An example of this visualization is shown in Figure \ref{fig:seq}.

\captionsetup[subfigure]{labelformat=empty}

A dataset composed of 19 DSA series from 10 patients was used to conduct a preliminary evaluation. A clinician provided ratings for the prepared visualizations and their expected impact on the clinical workflow. All visualizations were anticipated to decrease the required mental load and reduce the time needed for the analysis of the angiographic series.
This holds true despite 5.3\% of visualizations being categorized as somewhat misleading and significant errors being found in 10.5\% of series. In these cases, the color coding was reported to nevertheless provide valuable information about vascular flow. 
A further 10.5\% of visualizations received neutral scores, 47\% were reported to have minor errors, and 31.6\% were given a full score. 
Both the confidence and accuracy of clinical staff during diagnosis, treatment planning, or while judging the effect of treatment were expected to be impacted positively by over 94\% of visualizations respectively.

\section{CONCLUSION}
In this paper, we proposed a novel approach to identify vessels contributing to an AVM by processing DSA image series. This approach holds great promise in the field of cerebrovascular diagnosis and treatment planning. 
We have shown that a combination of supervised and unsupervised approaches performs well when evaluated on real patient data. 
The usage of this hybrid method circumvents the problem of scarcity of real labeled data.
Future work will consist on conducting a more extensive user study with a larger cohort of expert and non-expert neurovascular surgeons and neuroradiologist, in addition to to extending the application of this technique to high framerate \cite{Haouchine2021Estimation}, high-resolution DSA acquisition to improve its accuracy. 

\section{ACKNOWLEDGEMENT}
The authors were partially supported by the following National Institutes of Health grants: R03EB032050, R03EB033910, and R01EB034223.

\bibliography{main}   
\bibliographystyle{spiebib}   

\end{document}